\def\H        {{$^1$H \/}}
\def\HH       {{$^2$H \/}}
\def\C        {{$^{13}$C \/}}
\def\N        {{$^{14}$N \/}}
\def\NN       {{$^{15}$N \/}}
\def\F        {{$^{19}$F \/}}
\def\Si       {{$^{29}$Si \/}}
\def\P        {{$^{31}$P \/}}

\def\second   {{2$^{\rm nd}$ \/}}

\def\fourth   {{4$^{\rm th}$ \/}}
\def\eighth   {{8$^{\rm th}$ \/}}

\newcommand{\mr}[1]{\mathrm{#1}}
\newcommand{\unit}[1]{\,\mathrm{#1}}

\newcommand{\aperp}{a_\perp}
\newcommand{\apar}{a_{||}}

\newcommand{\fac}{f_\mr{ac}}
\newcommand{\wo}{\omega_0}
\newcommand{\wone}{\omega_1}

\newcommand{\waco}{\Omega}

\newcommand{\Brms}{B_\mr{ac}^\mr{rms}}
\newcommand{\Dw}{\Delta_z}
\newcommand{\tpi}{t_\pi}
\newcommand{\theff}{\theta_\mr{eff}}
\newcommand{\yn}{\gamma_n}
\newcommand{\ye}{\gamma_e}

\newcommand{\Sx}{\hat{S}_x}
\newcommand{\Sy}{\hat{S}_y}
\newcommand{\Sz}{\hat{S}_z}
\newcommand{\Ix}{\hat{I}_x}

\newcommand{\Iz}{\hat{I}_z}

\newcommand{\phipi}{\phi_\pi}
\newcommand{\phifree}{\phi_\mr{free}}

\newcommand{\captionstyle}{\normalfont} 

\documentclass[prx,aps,twocolumn]{revtex4}

\newcommand{\sectionvspace}{0.15cm}\newcommand{\figurewidth}{0.47\textwidth}

\usepackage{graphicx}
\usepackage{bm}
\usepackage{amsmath}
\usepackage{amssymb}
\usepackage{verbatim}
\usepackage[colorlinks=true, pdfstartview=FitV, linkcolor=blue, citecolor=blue, urlcolor=blue]{hyperref} 

\begin{document}

\global\emergencystretch = .1\hsize 

\title{Spurious harmonic response of multipulse quantum sensing sequences}

\author{M. Loretz$^1$, J. M. Boss$^1$, T. Rosskopf$^1$, H. J. Mamin$^2$, D. Rugar$^2$, and C. L. Degen$^1$}
  \email{degenc@ethz.ch} 
  \affiliation{
   $^1$Department of Physics, ETH Zurich, Otto Stern Weg 1, 8093 Zurich, Switzerland. \\
	 $^2$IBM Research Division, Almaden Research Center, 650 Harry Road, San Jose, CA 95120, USA.
	}
	

\begin{abstract}
Multipulse sequences based on Carr-Purcell decoupling are frequently used for narrow-band signal detection in single spin magnetometry.
We have analyzed the behavior of multipulse sensing sequences under real-world conditions, including finite pulse durations and the presence of detunings.  We find that these non-idealities introduce harmonics to the filter function, allowing additional frequencies to pass the filter.
In particular, we find that the XY family of sequences can generate signals at the $2\fac$, $4\fac$ and $8\fac$ harmonics and their odd subharmonics, where $\fac$ is the ac signal frequency.  Consideration of the harmonic response is especially important for diamond-based nuclear spin sensing where the NMR frequency is used to identify the nuclear spin species, as it leads to ambiguities when several isotopes are present.
\end{abstract}


\maketitle


\begin{center}{\bf I. INTRODUCTION}\end{center}

Multipulse decoupling sequences, initially developed in the field of nuclear magnetic resonance (NMR) spectroscopy \cite{carr54,slichter}, have enjoyed a renaissance for the control of individual quantum systems \cite{du09,ryan10,delange10,naydenov11}.  The concept relies on periodic reversals of the coherent evolution of the system, where the effect of the environment is canceled over the complete sequence.  Multipulse decoupling can be regarded as an efficient high-pass filter that averages out low frequency noise.

\begin{figure}[t!]
\centering
\includegraphics[width=\figurewidth]{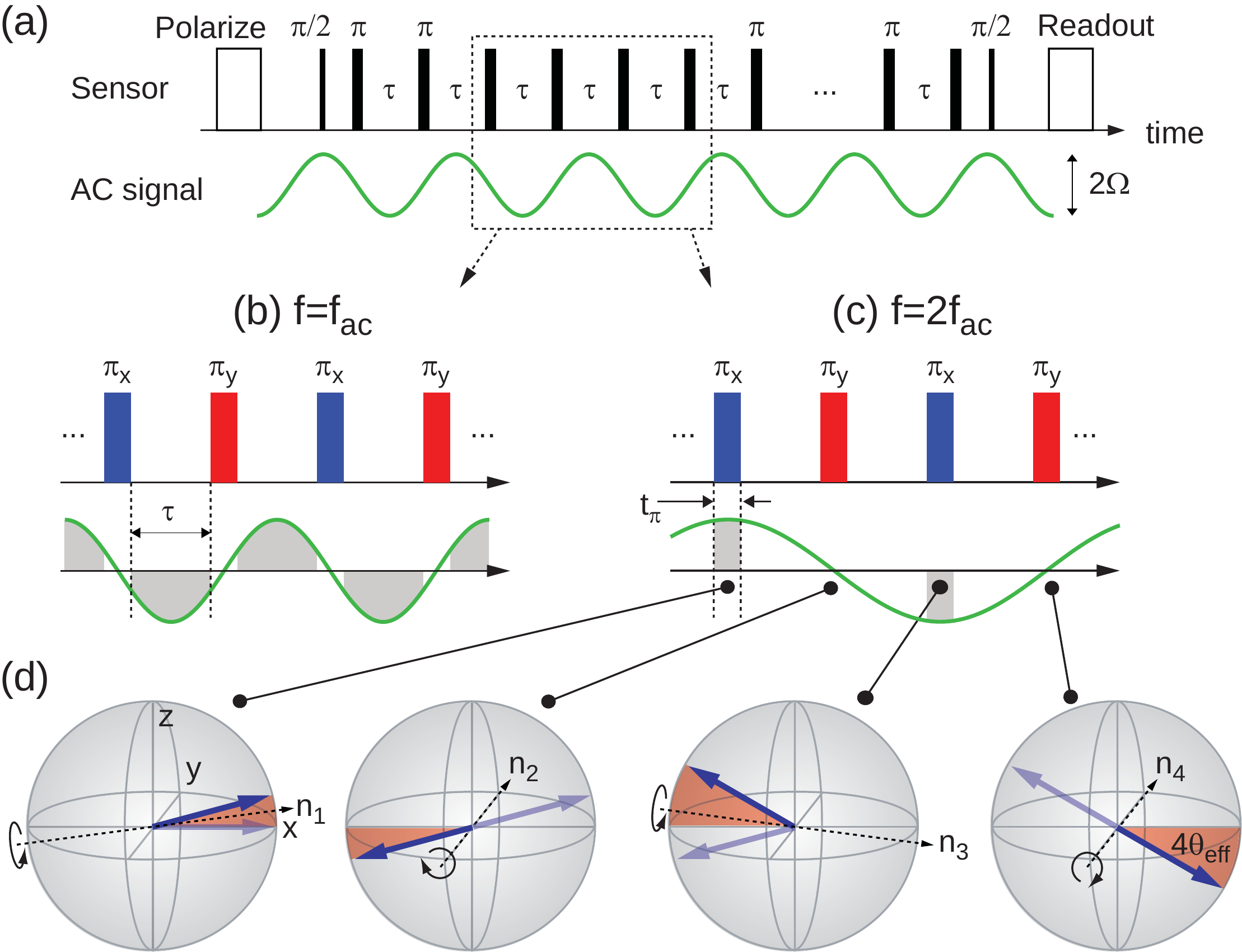}
\caption{\captionstyle
(a) Generic scheme of multipulse quantum sensing based on Carr-Purcell decoupling.  The detection frequency $f=1/(2\tau)$ is adjusted to match the ac signal frequency $\fac$.
(b,c) Sensor phase accumulation for the XY4 sequence for the fundamental signal $f=\fac$ and the \second harmonic signal $f=2\fac$.  Gray shaded areas show periods where phase accumulation occurs.
(d) Bloch vector picture of the basic mechanism for phase accumulation at the \second harmonic.
${\bf n_{1..4}}$ are rotation axes of $\pi$ pulses and $\theff$ is the angle of the effective field, as explained in the text.
}
\label{fig:bloch}
\end{figure}

It has recently been recognized that decoupling sequences with equal pulse spacing offer an opportunity for sensitive ac signal detection \cite{delange11,zhao11}.  By tuning the interpulse delay $\tau$, the sequence can be made commensurate with a signal's periodicity leading to recoupling -- that is, the decoupling fails for a specific set of signal frequencies that are an odd multiple of $1/(2\tau)$ (see Fig. \ref{fig:bloch}a).  The quantum system then effectively acts as a narrow-band lock-in amplifier \cite{kotler11} with demodulation frequency $f=k/(2\tau)$ and approximate bandwidth $f/(N/2)$, where $N$ is the total number of pulses in the sequence and $k=1,3,5,...$ is the resonance order.  A precise transfer function of decoupling filters has been given in several recent papers \cite{cywinski08,kotler11,delange11}.

Multipulse sensing can greatly improve detection sensitivity, as it selectively measures the influence of a desired ac signal while suppressing unwanted noise.  Moreover, the technique presents an opportunity to perform spectroscopy over a wide frequency range \cite{cywinski08,alvarez11,bylander11,gustavsson12,romach15}.  A particularly important application has been the detection of nanoscale NMR signals using single spins in diamond where the nuclear Larmor frequency is taken as a fingerprint for the detected spin species \cite{taminiau12,zhao12,kolkowitz12,shi14,staudacher13,loretz14,muller14,devience15,rugar15,haberle15,loretz14science}.  Although mostly developed in the context of single spin magnetometry, multipulse sensing sequences have been applied to other quantum systems including trapped ions \cite{kotler11} and superconducting qubits \cite{bylander11,gustavsson12}.  

In this paper we consider a specific class of multipulse sensing sequences known as the XY-family of sequences \cite{gullion90}.  XY-type sequences use the common template of equidistant $\pi$ pulses, but pulse phases are judiciously alternated so as to cancel out pulse imperfections, such as pulse amplitude or detuning.  The XY-family has become widely popular for measurements that require a large number of pulses $N$, and it has been the sequence of choice for most reported NV-based NMR experiments \cite{staudacher13,loretz14,muller14,devience15,rugar15,haberle15,loretz14science}.

Here we show that the phase alternation of XY-type sequences causes additional frequencies to pass the multipulse filter.  In particular, we show that an ac signal with frequency $\fac$ will produce response at the \second, \fourth and \eighth harmonics (and their $k$'th subharmonics), depending on the sequence used.  The harmonics are caused by a combination of time evolution during the finite duration of $\pi$ pulses and the ``superperiods'' introduced by phase cycling.  Although the feature is generic to all experiments, it is most prominent for low pulse amplitudes, short interpulse delays, and if a static detuning is present.  The feature is significant because it leads to further ambiguities in signal analysis and complicates interpretation of spectra.
%


\vspace{\sectionvspace}\begin{center}{\bf II. THEORY}\end{center}

In order to understand the generation of harmonics we consider the simple case of an XY4 sequence exposed to an ac signal with frequency $\fac$.  The basic building block of the XY4 sequence consists of four $\pi$ pulses with alternating $X$,$Y$ phases \cite{gullion90} (see Fig. \ref{fig:bloch}b,c).  In an ordinary sensing experiment, the interpulse spacing $\tau$ is matched to half the periodicity of the ac signal ($f=\fac$), leading to a constructive addition of phase during free evolution intervals.  The phase accumulated per period $\tau$ is
\begin{equation}
\phifree = \frac{2\waco\tau}{\pi} \ ,
\end{equation}
and the phase accumulated during the entire sequence is $N\phifree$, where $\waco$ is the amplitude of the ac signal.  $\waco$ represents a coupling constant with units of angular frequency.

In any realistic experimental implementation, $\pi$ pulses have a finite duration, and additional time evolution of the quantum system will occur \cite{pasini11}.  For the example of the XY4 sequence, we find that a feature of the additional time evolution is phase accumulation at the second harmonic where $f=2\fac$.  This phase accumulation is different from $\phifree$ in that it occurs during $\pi$ pulses, and not during free evolution intervals.

The feature can be understood by a picture of Bloch vector rotations (Fig. \ref{fig:bloch}d).  For simplicity, we assume that the ac field has maximum amplitude during $\pi_X$ pulses (and is zero during $\pi_Y$ pulses), and we neglect the free evolution during $\tau$.  The action of the XY4 block can then be described by a set of four $\pi$-rotations $R_{\bf\hat{n}_{1...4}}^{\pi}$ around the axes
\begin{eqnarray}
{\bf \hat{n}_1} & = &  (\cos\theff,0,\sin\theff) \ , \nonumber \\
{\bf \hat{n}_2} & = &  (0,1,0) \ , \nonumber \\
{\bf \hat{n}_3} & = &  (\cos\theff,0,-\sin\theff) \ , \nonumber \\
{\bf \hat{n}_4} & = &  (0,1,0) \ .
\end{eqnarray}
Axes ${\bf \hat{n}_1}$ and ${\bf \hat{n}_3}$ correspond to the $x$-axis tilted by the effective field angle $\theff = \tan^{-1}(\waco/\wone)\approx \waco/\wone$ \cite{slichter}, where $\wone$ is the angular velocity (Rabi frequency) of $\pi$ rotations.  The key feature here is that ${\bf \hat{n}_1} \not\parallel {\bf \hat{n}_3}$ due to the change in sign of the ac signal.  Assuming the Bloch vector is initially aligned with the $x$-axis, the vector orientation after the four $\pi$-rotations is
\begin{equation}
R_{\bf \hat{n}_4}^\pi R_{\bf \hat{n}_3}^\pi R_{\bf \hat{n}_2}^\pi R_{\bf \hat{n}_1}^\pi (1,0,0) = (\cos 4\theff,0,-\sin 4\theff) \ .
\end{equation}
This corresponds to a net rotation around the $y$-axis by an angle $4\theff \approx 4\waco/\wone$.  The sensor therefore acquires an additional ``anomalous'' phase during the XY4 block, on average
\begin{equation}
\phipi \approx \theff \approx \frac{\waco}{\wone} = \frac{\waco\tpi}{\pi}
\label{eq:phipi}
\end{equation}
per $\pi$-pulse, where $\tpi$ is duration of the square-shaped pulse.  A similar calculation can be made for other harmonics and pulse sequences, and resulting phases have been collected in Table \ref{table:phases}.  These values are quantitative for the evolution during a single XY-block in the limit of weak coupling, but only qualitative for longer sequences (due to backaction) or if a detuning is present.
\begin{table}[t!]
\centering
\begin{tabular}{lcccccc}
\hline\hline
Harmonic & $2$ & $4$ & $8$ & $2/3$ & $4/3$ & $8/3$ \\
\hline
CPMG	& ---					 & ---					& ---					 & ---					& ---					 & ---					 \\
XY4		& $1.00\theff$ & ---          & ---          & $1.00\theff$ & ---          & ---          \\
XY8   & $0.71\theff$ & $0.27\theff$ & ---          & $0.71\theff$ & $0.65\theff$ & ---          \\
XY16 	& $\ast$       & $\ast$       & $0.21\theff$ & $\ast$       & $\ast$    & $0.91\theff$ \\
\hline\hline
\ & $2/5$ & $4/5$    & $8/5$    & $2/7$    & $4/7$    & $8/7$ \\
\hline
CPMG	& ---					 & ---					& ---					 & ---					& ---					 & ---					\\
XY4		& $1.00\theff$ & ---          & ---          & $1.00\theff$ & ---          & ---          \\
XY8   & $0.71\theff$ & $0.65\theff$ & ---          & $0.71\theff$ & $0.27\theff$ & ---          \\
XY16 	& $\ast$       & $\ast$       & $0.18\theff$ & $\ast$       & $\ast$       & $0.32\theff$ \\
\hline\hline
\end{tabular}
\caption{Anomalous phase $\phipi$ at harmonics $f/\fac$ for CPMG, XY4, XY8 and XY16 sequences (see Ref. \cite{gullion90} for sequence definition).  Values were determined by a numerical simulation as described in Section IV.  $\theff\approx\waco/\wone$ is the angle of the effective field.  Entries marked by $\ast$ indicate that harmonics only appear in the presence of a static detuning.}
\label{table:phases}
\end{table}

The anomalous phase $\phipi$ can be compared to the ordinary phase $\phifree$, providing a ``relative strength'' $r = \phipi/\phifree$ of harmonics compared to the fundamental signal.  For the example of the \second harmonic in XY4 detection,
\begin{equation}
r = \frac{\phipi}{\phifree} = \frac{\tpi}{2\tau} \ .
\end{equation}
Other harmonics and sequences follow by using the appropriate multiplier from Table \ref{table:phases}.  We find that the phase accumulated during $\pi$ rotations is equal to the phase accumulated during free evolution, scaled by $\tpi/2\tau$.  Since most often $\tpi\ll\tau$, the harmonics will typically be much weaker than the fundamental signal.  The harmonics may, however, become relevant if one intends to detect a weak ac signal in the presence of a strong, undesired signal.  Note finally that higher order resonances ($k>1$) are not attenuated (sometimes enhanced) for anomalous signals, unlike the ordinary signals where $\phifree \propto 1/k$ \cite{taminiau12}.


\vspace{\sectionvspace}\begin{center}{\bf III. SINGLE-SPIN MAGNETOMETRY}\end{center}

The above considerations apply to several relevant situations in single spin magnetometry of nanoscale NMR signals.  Here, the precessing nuclear magnetization from single nuclei \cite{taminiau12,zhao12,kolkowitz12,loretz14science} or nuclear ensembles \cite{staudacher13,loretz14,muller14,devience15,rugar15,haberle15} provides the ac signal.  In the following we focus our attention on the specific case of an electron-nuclear two-spin system, where the electronic spin serves as the quantum sensor.  The Hamiltonian of this system in a rotating frame of reference is
\begin{equation}
\hat{H} = \underbrace{\Dw \Sz}_{\mr{static}}
+ \underbrace{\wone\{x_\mr{mod}(t)\Sx + y_\mr{mod}(t)\Sy\}}_{\mr{control}}
+ \underbrace{\aperp \Sz\Ix + \wo \Iz}_{\mr{ac\ signal}}
\label{eq:hamiltonian}
\end{equation}
where we have grouped terms into static, control and ac contributions.  $\Dw$ represents a static detuning of the electron spin $\hat{S}$ with respect to the rotating frame of reference, $x_\mr{mod}(t)$ and $y_\mr{mod}(t)$ represent the amplitude modulation of the multipulse decoupling sequence (with values between -1 and 1), $\aperp$ represents the transverse coupling to the nuclear spin $\hat{I}$, and $\wo = \yn B + \apar \Sz \approx \yn B + \apar/2$ is the effective Larmor frequency of the nuclear spin composed of static bias field $B$ and parallel hyperfine field contribution $\apar$ (see supplementary material to Ref. \cite{loretz14science} for a detailed discussion).  $\yn$ is the nuclear gyromagnetic ratio.

Returning to our generic expression for anomalous phase accumulation, Eq. (\ref{eq:phipi}), we can associate $\waco \rightarrow \aperp/2$ and $\fac \rightarrow \wo/(2\pi)$.  For the case of a large nuclear spin ensemble with rms nuclear field $\Brms$, one would associate $\waco \rightarrow \ye\Brms/2$, where $\ye$ is the electron gyromagnetic ratio.
\begin{figure}[t!]
\centering
\includegraphics[width=\figurewidth]{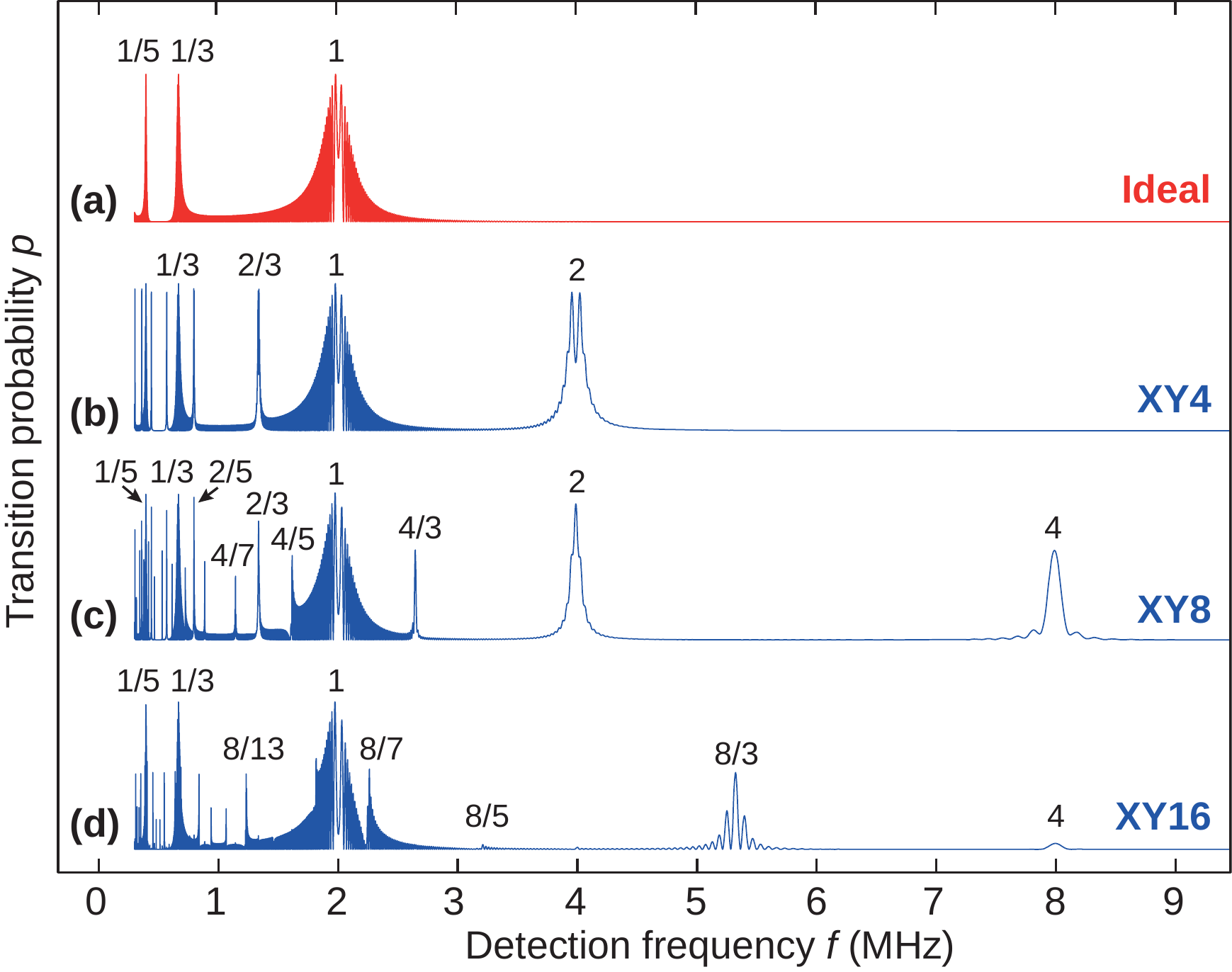}
\caption{\captionstyle
Simulated spectra showing response as a function of detection frequency $f=1/(2\tau)$ for coupling to a single nuclear spin with $\fac=\wo/(2\pi)=2\unit{MHz}$ and $\waco=\aperp/2=2\pi\cdot 200\unit{kHz}$.
(a) Ideal sensing sequence with infinitely short $\pi$ pulses.  Only the expected peaks at $\fac/k$ are observed.
(b-d) Real XY4, XY8 and XY16 sequences \cite{gullion90} with finite, square-shaped $\pi$ pulses.  Additional peaks at $2\fac/k$, $4\fac/k$ and $8\fac/k$ are observed, leading to a dense ``forest'' of peaks.
Number of pulses was $N=480$, Rabi frequency was $\wone/2\pi=20\unit{MHz}$, and detuning (for c,d) was $\Dw=1\unit{MHz}$.
}
\label{fig:filterfunction}
\end{figure}
%


\vspace{\sectionvspace}\begin{center}{\bf IV. SIMULATIONS}\end{center}

We have performed a set of numerical simulations \cite{loretz14science} to investigate the time evolution of the Hamiltonian, Eq. (\ref{eq:hamiltonian}).  Our quantity of interest was the probability that the spin state at the end of the sequence $|\alpha\rangle$ deviated from its original state $|0\rangle$, expressed by the transition probability $p=1-|\langle \alpha|0\rangle|^2$.

Fig. \ref{fig:filterfunction} presents simulated spectra for different multipulse sensing sequences.  The top panel shows the filter response for ideal $\pi$ rotations of infinitely short duration.  As expected, signal peaks are generated at frequencies $f = \fac/k$, where $k=1,3,5,...$.  Lower panels, by contrast, show the filter response of XY sequences for real $\pi$ rotations that have a finite duration.  Many extra peaks appear corresponding to \second, \fourth and \eighth harmonics of $\fac/k$.  The spectra in Fig. \ref{fig:filterfunction} represent the response to a single ac signal (single spin) with frequency $\fac$.  Obviously, if two or more signals were present, analysis of spectra would quickly become intractable.

\begin{figure}[t!]
\centering
\includegraphics[width=\figurewidth]{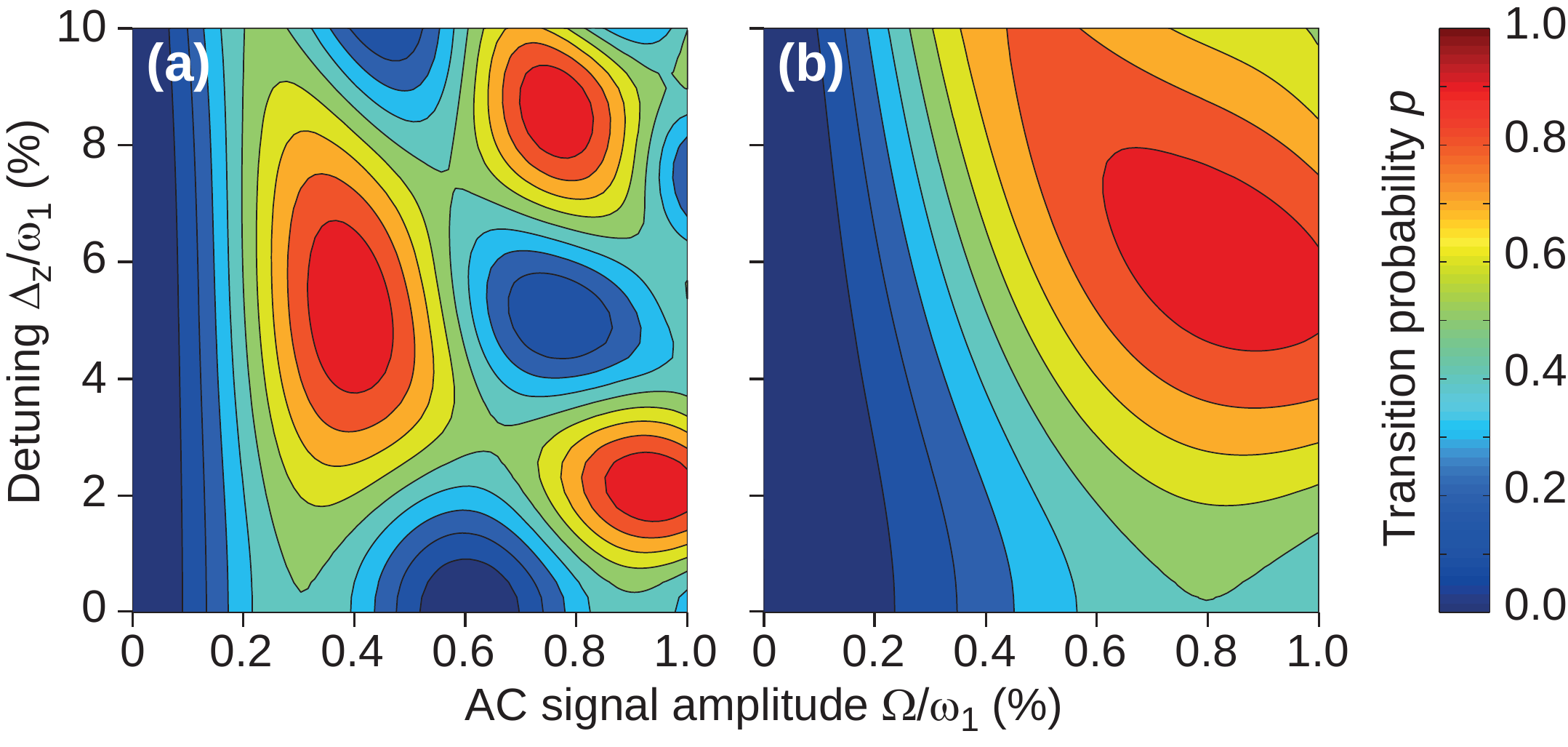}
\caption{\captionstyle
Simulated transition probability for the (a) \second and (b) \fourth harmonic of the XY8 sequence.
Horizontal axis is ac signal amplitude and vertical axis is detuning.
Further simulation parameters were $N=1,024$ and $\wone/2\pi=20\unit{MHz}$.
}
\label{fig:detuning}
\end{figure}
Fig. \ref{fig:detuning} investigates the influence of a static detuning $\Dw$.  Several effects may be noticed.  First, a static detuning exacerbates the harmonic peaks -- stronger anomalous response is generated for the same ac signal magnitude $\waco$.  Second, although not evident from this plot, peaks appear even at those harmonics where $\phipi = 0$ in Table \ref{table:phases}.  Third, the time evolution of $p$ becomes markedly different -- values oscillate between $p=0...0.5$ in the absence of a detuning, whereas they can oscillate between $p=0...1$ and become aperiodic in the presence of a detuning.


\vspace{\sectionvspace}\begin{center}{\bf V. EXPERIMENTS}\end{center}

We have experimentally verified the existence of harmonics using the nitrogen-vacancy (NV) center in diamond as the single spin sensor.  The NV center is a prototype electron spin system ($S=1$) that can be optically detected at room temperature \cite{jelezko06} and that has served as a testbed for many recent multipulse sensing experiments.  For our measurements, the NV center was polarized and read out using non-resonant green laser excitation, and manipulated using microwave control pulses with adjustable phase and amplitude \cite{loretz14}.  A static bias field was applied to lift the spin degeneracy and all sensing experiments were carried out on the $m_S=0\leftrightarrow m_S=-1$ subsystem.

\begin{figure}[t!]
\centering
\includegraphics[width=\figurewidth]{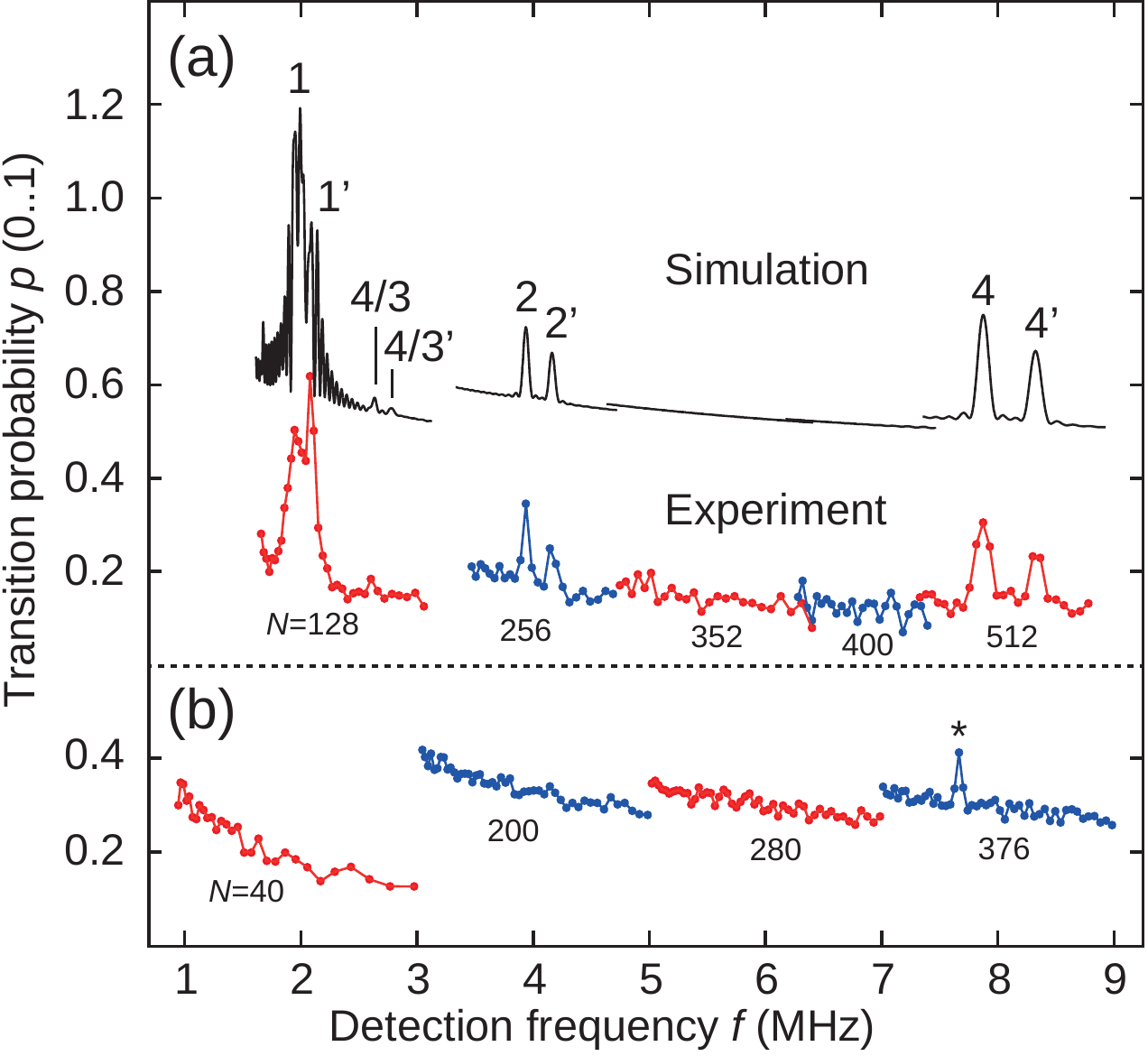}
\caption{\captionstyle
(a) Spectrum recorded by an NV center with two weakly coupled \C nuclei, measured using an XY8 sequence.
Multiple harmonics are clearly observed.  Colors represent different datasets.  Coupling constants for the two \C were $\aperp/2\pi=220\unit{kHz}$ and $\aperp'/2\pi=180\unit{kHz}$.
The number of pulses $N$ is given in the plot.  The associated total evolution time was $T=N/(2f)$. 
Further parameters were $\wone/2\pi=28\unit{MHz}$ and $\Dw=1.5\unit{MHz}$ (due to the \NN hyperfine splitting).
(b) For comparison, spectrum recorded by an NV center with no \C nearby.  Peak marked by $\ast$ is due to an ensemble of \H with $\Brms=400\unit{nT}$.
}
\label{fig:13Cspectrum}
\end{figure}
In a first experiment we have recorded the XY8 response of an NV center with two proximal \C nuclei ($I=1/2$) in a bias field of $183\unit{mT}$.  The nuclear Larmor frequencies of the two \C nuclei in this field were $\fac = 1.97\unit{MHz}$ and $\fac' = 2.08\unit{MHz}$. (The frequencies were slightly different due to a small parallel hyperfine contribution.)  Fig. \ref{fig:13Cspectrum} shows the spectral response over the frequency range of $f=1.5-9\unit{MHz}$.  We found that almost all harmonics in that range could be resolved and matched with simulations using a single set of parameters.

\begin{figure}[t!]
\centering
\includegraphics[width=\figurewidth]{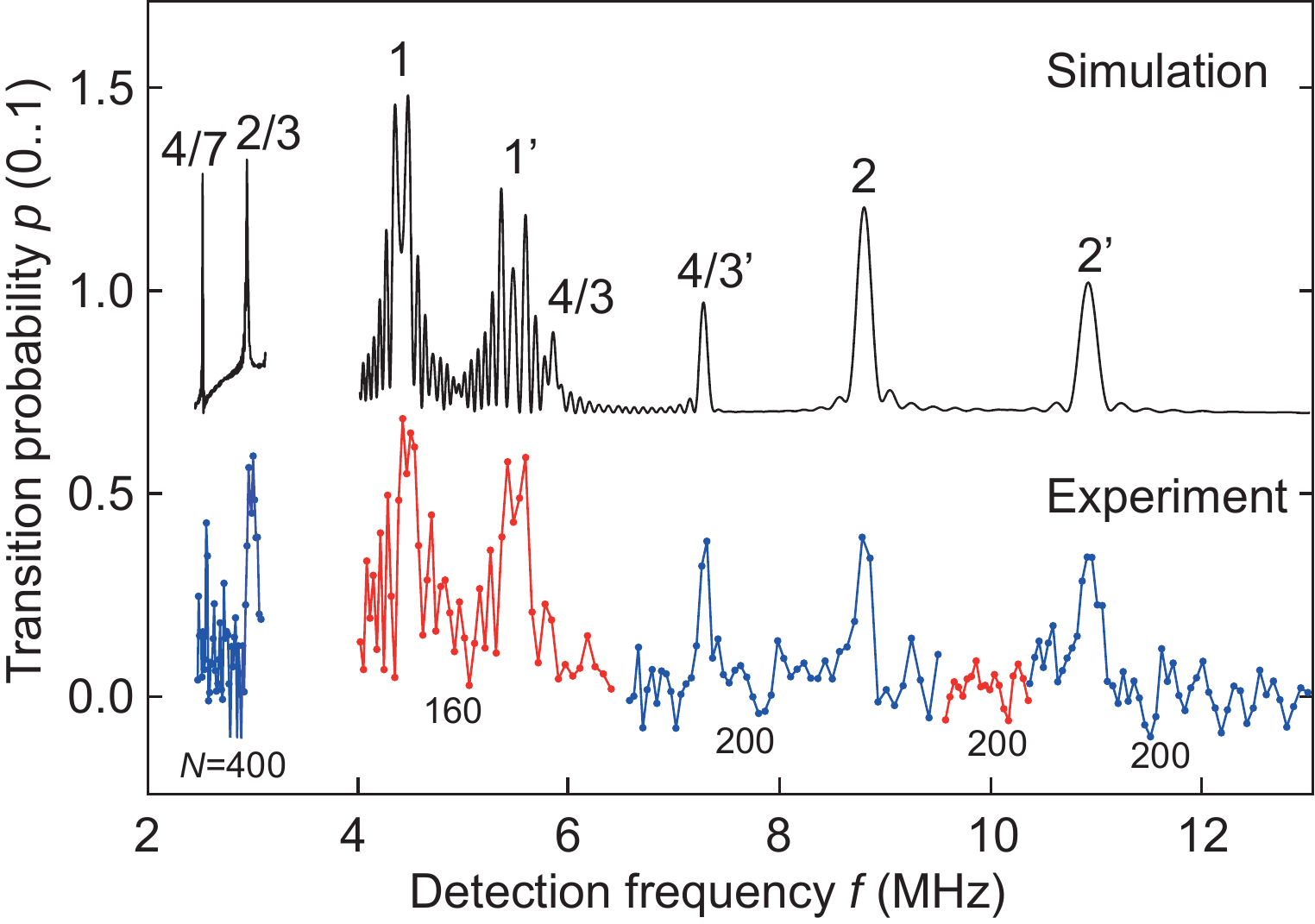}
\caption{\captionstyle
Spectrum recorded by an NV center coupled to its own \N nucleus, measured using an XY8 sequence.
Common coupling constant was $\aperp/2\pi=280\unit{kHz}$.
This coupling was due to a deliberate misalignment of the bias field of $\sim 8^\circ$ with respect to the NV symmetry axis.
Further experimental parameters were similar to Fig. \ref{fig:13Cspectrum}.
}
\label{fig:14Nspectrum}
\end{figure}
Fig. \ref{fig:14Nspectrum} shows a second example where the NV center's electron spin was coupled to its own \N nucleus ($I=1$).  The \N has two nuclear spin transitions, resulting in two signals with frequencies $\fac = 4.4\unit{MHz}$ and $\fac' = 5.4\unit{MHz}$.  Again, most expected harmonics could be observed and matched with simulations.


\vspace{\sectionvspace}\begin{center}{\bf VI. AMBIGUITIES BETWEEN NMR SIGNALS}\end{center}

The feature of harmonics is of particular significance for recent nanoscale NMR experiments with near-surface NV centers.  These experiments used spectral identification to discriminate different nuclear isotopes in samples by their NMR frequencies, most prominently \H, \C, \F, \Si and \P \cite{staudacher13,loretz14,muller14,devience15,rugar15,haberle15}.

\begin{figure}[b!]
\centering
\includegraphics[width=\figurewidth]{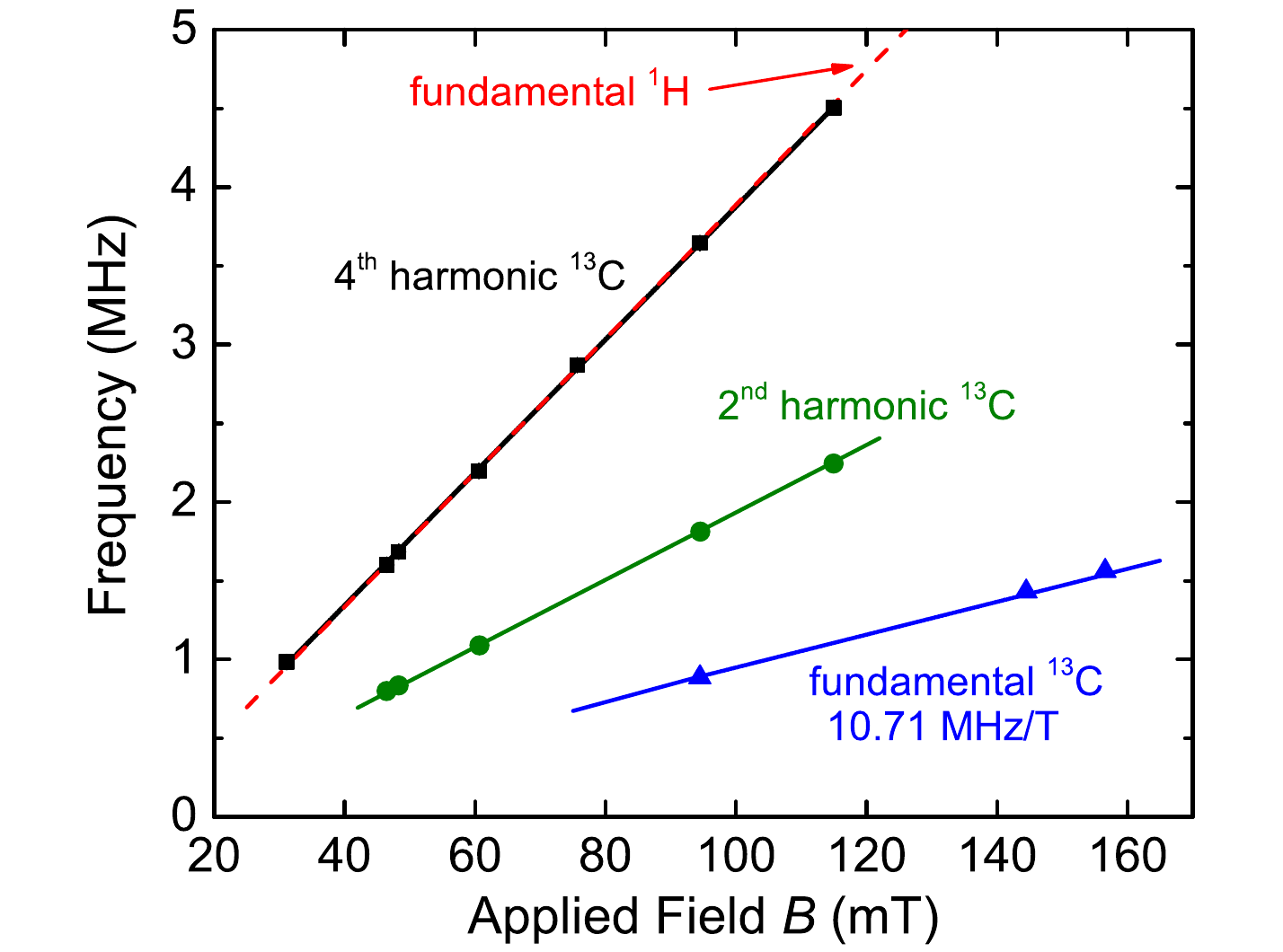}
\caption{\captionstyle
Frequency of peak response for various magnetic bias fields $B$, measured using an XY8 sequence.
Points are measured peak positions. Solid lines have slopes chosen to be $1\times$, $2\times$ and 4$\times$ the gyromagnetic ratio of \C. The red dashed line
has slope given by the gyromagnetic ratio of \H, and is almost identical to the \C \fourth harmonic line (black line).
}
\label{fig:13Clarmor}
\end{figure}
We now show that in the presence of more than one nuclear species, harmonics can produce coincidential overlap between signals and lead to ambiguities in peak assignments.  
As a particular example we consider the detection of a weak \H signal in the presence of \C, which is a situation typical to NV centers in natural abundance diamond substrates ($\sim1$\% \C content) \cite{loretz14science}.  Here the signal overlap arises because the \H NMR frequency coincides with the $4\times$ harmonic of \C to within 0.6\%.  We find that single \C nuclei can generate signatures virtually identical to those of single \H, including a Zeeman scaling with magnetic field and a quantum-coherent oscillation of the signal.

We begin with the scaling of the peak frequency with bias field $B$, as shown in Fig. \ref{fig:13Clarmor}. The measured peak positions (points) can all be associated with the response from a single \C nucleus at either its fundamental Larmor frequency, or its \second or \fourth harmonic. The linear slope of the fundamental frequency (blue line) is given by the gyromagnetic ratio of \C ($\yn = 10.71\unit{MHz/T}$) and is indicative of the nuclear species, while the slopes of the two harmonic frequencies are enhanced with apparent gyromagnetic ratios of $2\yn$ and $4\yn$.  Since the value of $4\yn = 42.82\unit{MHz/T}$ coincides with the gyromagnetic ratio of protons ($\yn = 42.57\unit{MHz/T}$) within experimental uncertainty, the nuclear isotope cannot be uniquely identified.
\begin{figure}[t!]
\centering
\includegraphics[width=\figurewidth]{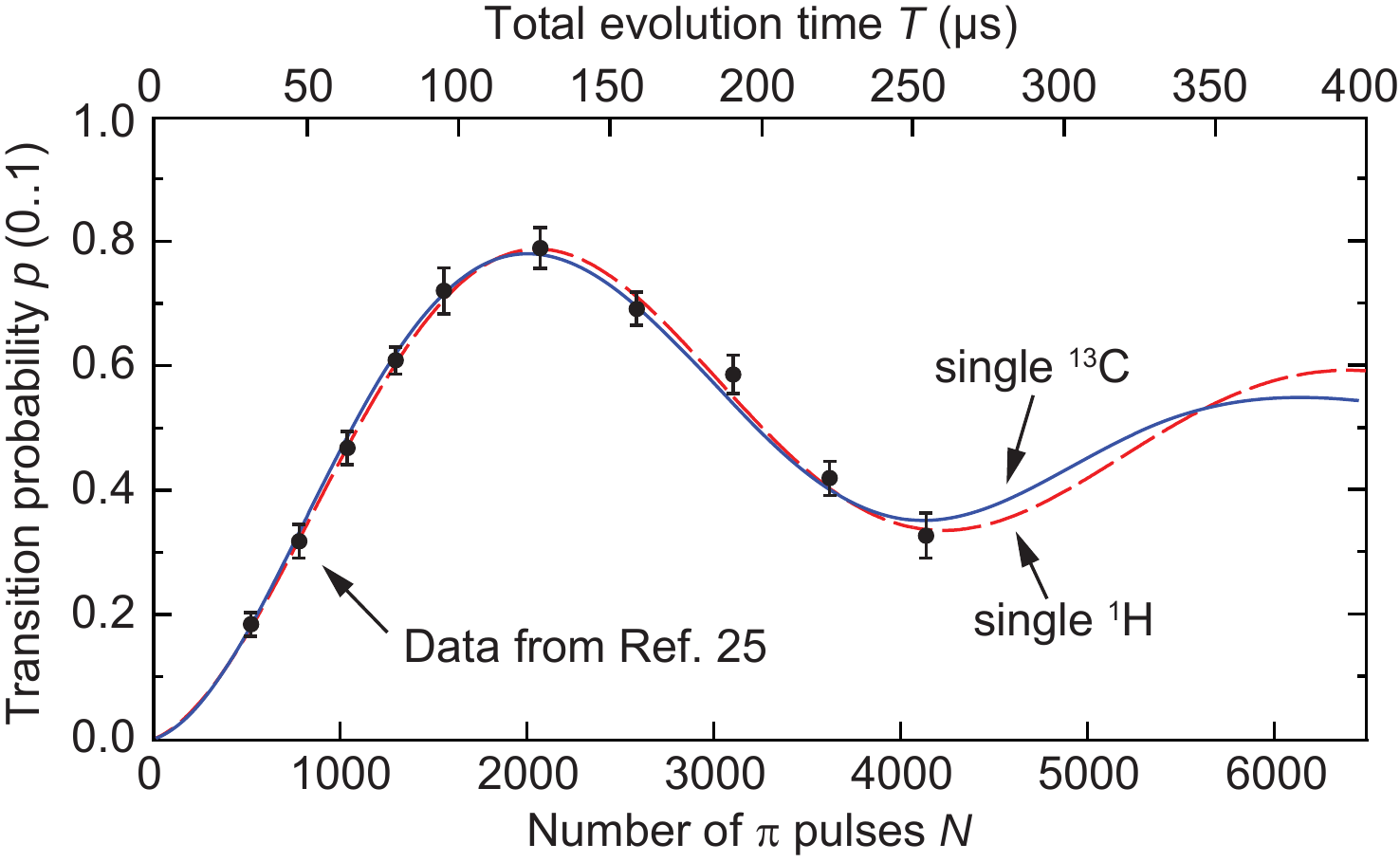}
\caption{\captionstyle
Peak transition probability with increasing total evolution time $T=N\tau$ measured at the expected \H resonance.
Black data points are taken from Ref. \cite{loretz14science}.
Red dashed curve shows simulation for a single \H spin with coupling constant $\aperp/2\pi = 14\unit{kHz}$.
Blue solid curve shows simulation for the \fourth harmonic of a single \C spin with a coupling constant $\aperp/2\pi = 150\unit{kHz}$.
Pulse sequence was XY8.
Further parameters were $B=182\unit{mT}$, $\wone/2\pi=25\unit{MHz}$ and $\tau=61.5\unit{ns}$.
}
\label{fig:timeevolution}
\end{figure}
\begin{table}[b!]
\centering
\begin{tabular}{cccccc}
\hline\hline
Isotope & Frequency & Mimicking & Harmonic & Harmonic	 & Rel. freq. \\
\ 		  & \ 				& isotope   & \        & frequency & difference \\
\hline
\H		&	4.257 MHz		& \C  & $4\times$    &	4.282 MHz		& 0.6\% \\ 
\HH		&	0.654 MHz		& \H  & $2/13\times$ &	0.655 MHz		& 0.2\% \\ 
\Si		&	0.846 MHz		& \H  & $1/5\times^\ast$ &	0.851 MHz		& 0.7\% \\ 
\ 		&	\ 					& \C  & $4/5\times$  &	0.856 MHz		& 1.2\% \\ 
\P		&	1.723 MHz		& \H  & $2/5\times$  &	1.703 MHz		& 1.2\% \\ 
\hline\hline
\end{tabular}
\caption{Examples of coincidential overlap between fundamental and harmonic frequencies for important NMR isotopes. Values are given for a $100\unit{mT}$ bias field. All above harmonics are visible in Fig. \ref{fig:filterfunction}. $^\ast$Ordinary (not anomalous) harmonic. }
\label{table:overlap}
\end{table}
We found also the time evolutions of fundamental and harmonic signals produce indistinguishable signatures.  Fig. \ref{fig:timeevolution} shows the peak height of an apparent proton signal as the number of pulses $N$ in increased.  An oscillating signal is observed that can be reproduced by either of two simulations, one assuming the presence of a single \H and one of a single \C.  Again, no conclusion can be made on the identity of the nuclear species.

The above example illustrates that assignment of a signal based on a single peak is in general insufficient.  A more trustworthy identification could be obtained by measuring a wide spectral region containing several harmonics, as shown in Fig. \ref{fig:13Cspectrum}.  For the specific situation of \C and \H, the second harmonic at twice the \C Larmor frequency (roughly half the \H frequency) could serve as a tell-tale signature of \C.  This peak is absent for \H.

We note that the ambiguity between \H and \C is just one particular case.
Given the large number of harmonics and nuclear isotopes, many other pairs of nuclear species with potential overlap can be expected.
Table \ref{table:overlap} collects a few additional cases of particular relevance to NMR where such coincidential signal overlap could occur.
Consideration of harmonics is especially important in the presence of \H or for \C-containing diamond, because these species often produce a relatively strong signal.


\vspace{\sectionvspace}\begin{center}{\bf VII. SUMMARY}\end{center}

In summary, we have investigated the spectral filtering characteristics of an important class of multipulse quantum sensing sequences, the XY-family of sequences.  We found the time evolution during finite $\pi$ pulses, which is present in any experimental implementation, to cause phase build-up at higher harmonics of the signal frequency, leading to sets of additional peaks in the spectrum.  We have further investigated this feature by simulations and experiments of single NV centers in diamond.  The feature has particular significance for nanoscale NMR experiments that rely on spectral identification.  Specifically, we found that signal detection at a single frequency is in general insufficient to uniquely assign a certain nuclear spin species.

We finally mention several sensing schemes that do not suffer from the ambiguities inherent to multipulse sequences.  Namely these include rotating-frame spectroscopy \cite{yan13,loretz13,london13} and free precession techniques \cite{mamin13,laraoui13,taminiau14}.  Moreover, it may be possible to craft varieties of sequences that avoid harmonic resonances yet maintain superior static error compensation, such as sequences with composite $\pi$ pulses \cite{ryan10,souza11}, locally non-uniform pulse spacing \cite{zhao14}, or aperiodic phase alternation.  Although the alternative schemes have their own restrictions and may not always be available, they could offer independent verification of spectral features.


\vspace{\sectionvspace}\begin{center}{\bf ACKNOWLEDGMENTS}\end{center}

This work was supported by Swiss NSF Project Grant $200021\_137520$, the NCCR QSIT, the DIADEMS programme of the European Commission, and the DARPA QuASAR program.

\end{document}